\documentclass[twocolumn,dvipdfmx,showpacs,preprintnumbers,amsmath,amssymb,aps,prc,nofootinbib,superscriptaddress,eqsecnum]{revtex4}
\usepackage{amsmath}
\usepackage{braket}
\usepackage{graphicx}
\usepackage{verbatim}
\usepackage{setspace}
\usepackage{feynmf}
\usepackage{color}
\usepackage{booktabs}
\usepackage{multirow}
\usepackage{fancybox}
\usepackage{comment}

\newcommand{\lag}{{\cal L}}

\newcommand{\bea}[1]{\begin{align}#1\end{align}}

\newcommand{\braa}[1]{\left(#1\right)}

\newcommand{\brac}[1]{\left[#1\right]}

\newcommand{\tr}{{\rm tr}}

\begin{document}

\author{Yuichi Motohiro}
\affiliation{Department of Physics, Nagoya University, Nagoya, 464-8602, Japan}

\author{Youngman Kim}
\affiliation{Rare Isotope Science Project, Institute for Basic Science, Daejeon 305-811, Korea}

\author{Masayasu Harada}
\affiliation{Department of Physics, Nagoya University, Nagoya, 464-8602, Japan}

\title{Asymmetric nuclear matter in a parity doublet model with hidden local symmetry}

\begin{abstract}
We construct a model to describe dense hadronic matter at zero and finite temperature,
based on the parity doublet model of DeTar and Kunihiro, with including the iso-singlet scalar meson $\sigma$ as well as  $\rho$ and $\omega$ mesons.
We show that, by including a six-point interaction of $\sigma$ meson, the model reasonably reproduces the properties of the normal nuclear matter
with the chiral invariant nucleon mass $m_0$ in the range from $500~{\rm MeV}$ to $900~{\rm MeV}$.
Furthermore, we study the phase diagram based on the model, which shows that the value of the chiral condensate drops at the liquid-gas phase transition
point and at the chiral phase transition point.
We also study  asymmetric nuclear matter and find that the first order phase transition for the liquid-gas phase transition disappears in
asymmetric matter
and that the critical density for the chiral phase transition at non-zero density becomes smaller for larger asymmetry.
\end{abstract}


\maketitle

%%%%%%%%%%%%%%%%%%%%%%%%%%%%%%%%%%%%%%%%%%%%%%%%%%%%%%%%%%%%%%%%%%%%%%%%%%
\section{Introduction}

With the advent of next generation radioactive beam facilities
isospin asymmetric nuclear matter claims much attention in contemporary nuclear physics.
At those facilities we could create terrestrial environment to study dense matter with a large neutron
or proton excess through nuclear reactions with radioactive nuclei.

Studying nuclear matter is also
important to understand the structure   of neutron stars~\cite{Akmal:1998cf}.
 In 2010 and 2013,
two neutron stars with twice solar mass were found~\cite{Demorest:2010bx,Antoniadis:2013pzd} and many models yielding the soft equation of states (EOS) were excluded.
Neutron stars offer very cold and asymmetric dense environment and may have hyperons in the core of the stars.
If there are hyperonic degrees of freedom, it is expected that the EOS becomes softer and neutron star mass becomes lighter.
Another important astrophysical site for nuclear matter is a hybrid star whose center has quark matter \cite{Alford:2004pf}.

The properties of asymmetric matter have been investigated
in various approaches~\cite{Bombaci:1991zz, Muller:1995ji, Li:1997rc, Hofmann:2000vz, Liu:2001iz, Zuo:2001bd, vanDalen:2005ns, Chen:2007ih, Gogelein:2008vw,
Drews:2014spa}. Very recently liquid-gas and chiral phase transition are studied in a parity doublet model with a six-point scalar interaction in which
mesonic fluctuations are included by means of the functional renormalization group~\cite{Weyrich:2015hha}.
In this work we study isospin asymmetric dense matter in the framework of the parity doublet model (mirror assignment)~\cite{Detar:1988kn,jidookahosaka}.
The properties of symmetric dense matter such as chiral phase transition were extensively studied in the parity doublet models at zero
 or finite temperature~\cite{Hatsuda:1988mv, Zschiesche:2006zj, Sasaki:2010bp, Gallas:2011qp, Benic:2015pia}.
We extend the parity doublet model by including
$\rho$ and $\omega$ mesons through the
 hidden local symmetry and also by adding a six-point interaction of a scalar meson.
Here, as a first step, we will not consider hyperonic matter and work within the mean field approximation.

We determine our model parameters, except the chiral invariant mass ($m_0$),
by performing global fitting to physical inputs (masses and pion decay constant in free space and nuclear matter properties).
We then study the equation of state and the phase diagram of dense matter at finite temperature.
We find that the predicted slope parameter at the saturation density meets the constraint from heavy ion experiments and neutron star observations
(see, e.g. Refs.~\cite{lattimer_lim, slopeparametervalue})
 and observe that the chiral condensate drops at the chiral and liquid-gas transition
 points. It is also seen that smaller $m_0$ values prefer smaller critical densities for chiral phase transition.
The study of asymmetric matter reveals that the first order nature of the liquid-gas transition disappears in asymmetric matter and
the critical densities for the chiral transition become smaller with increasing asymmetries, which are consistent with previous studies.

In section \ref{model} we extend the parity doublet model, and  in section \ref{parameterF} we fix the model parameters.
Our results on bulk properties of nuclear matter and density dependence of chiral condensate and nucleon mass are given in section \ref{EoS}.
We present the phase diagram of dense (asymmetric) matter in section \ref{phaseD}.
Finally, conclusion and discussion follow in section \ref{SD}
%%%%%%%%%%%%%%%%%%%%%%%%%%%%%%%%%%%%%%%%%%%%%%%%%%%%%%%%%%%%%%%%%%%%%%%%%%
\section{Extended parity doublet model}\label{model}

We construct a chiral effective model based on the parity doublet model~\cite{Detar:1988kn, jidookahosaka},
in which a nucleon with positive parity is regarded as a chiral partner to the one with negative parity and they belong to the same multiplet.
The transformation properties of the positive and negative parity nucleon fields are given by
\bea{
\psi_{1r}&\to g_{R}\psi_{1r},\hspace{10pt}\psi_{1l}\to g_{L}\psi_{1l}\\
\psi_{2r}&\to g_{L}\psi_{2r},\hspace{10pt}\psi_{2l}\to g_{R}\psi_{2l}\ ,
}
where $g_{R}$ ($g_{L}$) is an element of SU(2)$_{R}$ (SU(2)$_{L}$) chiral symmetry group, and $\psi_{1r}$ and $\psi_{2r}$ ($\psi_{1l}$ and $\psi_{2l}$) are
the right-handed (left-handed) fields projected as
\begin{eqnarray}
&&
\psi_{1r,2r} = P_R \,\psi_{1,2} \ , \quad
\psi_{1l,2l} = P_L \, \psi_{1,2} \ , \quad
\nonumber\\
&& \qquad\qquad
P_{R,L} = \frac{1\pm \gamma_5}{2} \ .
\end{eqnarray}
To construct a linear sigma model including these nucleon fields, we introduce the following field:
\begin{equation}
M = \sigma+i{\vec\pi}\cdot{\vec\tau}\ ,
\end{equation}
where $\sigma$ denotes an iso-singlet scalar field, $\vec{\pi}$ the pion field, and
 ${\vec\tau}$ the Pauli matrices.
The chiral transformation property of  $M$ is given by
\bea{M\to g_{L} M g_{R}^\dag \ .}
By using these fields, the Lagrangian of a linear sigma model type is expressed as
\footnote{
This Lagrangian is rewritten into a more familiar form in the literatures as
\begin{align}
{\mathcal L}_{N}=&{\psi_{1}}i\gamma^\mu\partial_\mu\psi_1
+{\bar\psi_2}i\gamma^\mu\partial_\mu\psi_2
-m_0\left({\bar\psi_{1}}\gamma_5\psi_{2}-{\bar\psi_{2}}\gamma_5\psi_{1}\right)\notag\\&
-g_1{\bar\psi_{1}}\braa{\sigma+i\gamma_5{\vec\pi}\cdot{\vec\tau}}\psi_{1}
-g_2{\bar\psi_{2}}\braa{\sigma-i\gamma_5{\vec\pi}\cdot{\vec\tau}}\psi_{2}\label{pdmn}
\ \notag.
\end{align}
}
\begin{align}
{\mathcal L}_{N}=&{\bar\psi_{1r}}i\gamma^\mu{{D}_\mu}\psi_{1r}
+{\bar\psi_{1l}}i\gamma^\mu{{D}_\mu}\psi_{1l}\notag\\
&
+{\bar\psi_{2r}}i\gamma^\mu{{D}_\mu}\psi_{2r}
+{\bar\psi_{2l}}i\gamma^\mu{{D}_\mu}\psi_{2l}\notag\\
&-m_0\left[{\bar\psi_{1l}}\psi_{2r}-{\bar\psi_{1r}}\psi_{2l}
-{\bar\psi_{2l}}\psi_{1r}+{\bar\psi_{2r}}\psi_{1l}\right]\notag\\
&-g_1\left[{\bar\psi_{1r}} M^\dag\psi_{1l}
+{\bar\psi_{1l}} M \psi_{1r}\right]\notag\\
&
-g_2\left[{\bar\psi_{2r}} M\psi_{2l}
+{\bar\psi_{2l}} M^\dag \psi_{2r}\right] \ ,
\end{align}
where $m_0$ is the chiral invariant mass, $g_1$ and $g_2$ are the coupling constants,
and the covariant derivatives include the external gauge fields ${\mathcal R}_\mu$ and ${\mathcal L}_\mu$  as
\begin{align}
D_\mu \psi_{1r,2l} = \left( \partial_\mu - i {\mathcal R}_\mu \right) \psi_{1r,2l} \ ,
\notag\\
D_\mu \psi_{1l,2r} = \left( \partial_\mu - i {\mathcal L}_\mu \right) \psi_{1l,2r} \ .
\end{align}
The meson part of the Lagrangian is given by
\bea{\lag_{M}=&\frac{1}{4}\tr\left[\partial_\mu M \partial^\mu M^\dag\right] -V_{M}-V_{SB}\ ,
\label{meson lag}
}
where the meson potential $V_M$ and the explicit chiral symmetry breaking potential $V_{\rm SB}$ are
\begin{align}
V_{M}=&-\frac{1}{4}{\bar \mu^2}\tr\left[MM^\dag\right]+\frac{1}{16}\lambda\left\{\tr\left[MM^\dag\right]\right\}^2\notag\\
&
- \frac{1}{48} \lambda_6 \left\{ \mbox{tr} \left[ M M^\dag \right] \right\}^3
\ ,
\\
V_{\rm SB}=&-\frac{1}{4}\varepsilon\left(\tr\left[ {\mathcal M}^\dag M \right] +\tr \left[ {\mathcal M}  M^\dag\right] \right) \ .
\label{lsmm}
\end{align}
Here ${\mathcal M}$ is the quark mass matrix given as
\begin{align}
{\mathcal M} = \left( \begin{array}{cc} m_u & 0 \\ 0 & m_d \\ \end{array} \right)
\end{align}
and $\varepsilon$ is a constant of mass dimension two.
In the present analysis, we neglect the isospin breaking effect due to the mass difference of up and down quarks and take $m_u = m_d = \bar{m}$.
It should be noticed that the potential $V_M$ includes the dimension-six term which will play a very
important role in reproducing the properties of the normal nuclear matter with rather wide range of the chiral invariant nucleon masses. (See next section.)
When $\lambda_6 > 0$, the potential $V_M + V_{\rm SB}$ is not bounded from below.
In the present analysis, we determine the vacuum in the following way:
We first solve the stationary condition of the potential.  When there are more than one solutions, we choose the one with the lowest energy at the stationary point.

We next include $\rho$ and $\omega$ mesons into the model based on the hidden local symmetry (HLS) theory~\cite{hls, haradayamawaki}.
The HLS is introduced by performing the polar decomposition of the field $M$ as
\bea{M=\xi_L\sigma\xi_R=\sigma\xi_L^\dag\xi_R = \sigma U}
where $\sigma$ is a scalar meson field and $\xi_L$ and $\xi_R$
transform as
\bea{\xi_{L,R}\to h_{\omega}h_{\rho}\xi_{L,R}g_{L,R}^\dag
}
with $h_{\omega}\in\mbox{U(1)}_{\rm HLS}$ and $h_{\rho}\in\mbox{SU(2)}_{\rm HLS}$.
In the unitary gauge, $\xi_R$ and $\xi_L$ are parameterized as
\bea{\xi_R=\xi^\dag_L&=\exp(i\pi^aT^a/f_\pi)\ ,
}
where $T^a = \tau_a/2$ ($a=1,2,3$) with $\tau_a$ being the Pauli matrix.
In the HLS, the vector mesons are introduced as the gauge bosons of the HLS which transform as
\bea{\omega_{\mu} &\to h_{\omega}\cdot \omega_{\mu}\cdot h_{\omega}^\dag+\frac{i}{g_{\omega}}\partial_{\mu}h_{\omega}\cdot h_{\omega}^\dag \\
\rho_{\mu} &\to h_{\rho}\cdot \rho_{\mu}\cdot h_{\rho}^\dag+\frac{i}{g_{\rho}}\partial_{\mu}h_{\rho}\cdot h_{\rho}^\dag
\ ,
}
where $g_\omega$ and $g_\rho$ are the corresponding gauge coupling constants.

To construct a model Lagrangian with the HLS, it is convenient to introduce the following one-forms:
\begin{align}
{\hat\alpha_\perp^\mu} &\equiv\frac{1}{2i}\left
[D^\mu\xi_R\cdot\xi_R^\dag-D^\mu\xi_L\cdot\xi_L^\dag\right] \ ,
\notag\\
{\hat\alpha_\parallel^\mu} &\equiv\frac{1}{2i}\left[D^\mu\xi_R\cdot\xi_R^\dag+D^\mu\xi_L\cdot\xi_L^\dag\right]\ ,
\end{align}
where the covariant derivatives are given as
\begin{align}
D^\mu\xi_{L} &= \partial^\mu\xi_{L}+ig_{\rho}\rho^\mu\xi_{L}
 +ig_\omega \omega^\mu\xi_{L}+i\xi_L{\cal L}^\mu\ ,
\notag\\
D^\mu\xi_{R} & = \partial^\mu\xi_{R}+ig_\rho \rho^\mu\xi_{R}
  +i g_\omega \omega^\mu\xi_{R}+i\xi_R{\cal R}^\mu\ .
\end{align}
Now, the mesonic part of the Lagrangian extended by the HLS is expressed as
\bea{\lag_{M}=&\frac{1}{2}\partial_\mu\sigma\partial^\mu\sigma+\sigma^2\tr\left[{\hat\alpha_{\perp\mu}}{\hat\alpha_\perp^\mu}\right]-V_\sigma-V_{SB}\notag\\
&+\frac{m_{\rho}^2}{g_\rho^2}\tr\left[{\hat\alpha_{\parallel\mu}}{\hat\alpha_\parallel^\mu}\right]
+\left(\frac{m_\omega^2}{2g_\omega^2}-\frac{m_\rho^2}{2g_\rho^2}\right)
\tr\left[{\hat\alpha_{\parallel\mu}}\right]
\tr\left[{\hat\alpha_\parallel^\mu}\right]\notag\\
&-\frac{1}{2g_\rho^2}\tr\left[\rho_{\mu\nu}\rho^{\mu\nu}\right]
-\left(\frac{1}{2g_\omega^2}-\frac{1}{2g_\rho^2}\right)\tr\left[\omega_{\mu\nu}\right]\tr\left[\omega^{\mu\nu}\right]
\ ,
}
where the first line is from Eq.~(\ref{meson lag}) with the following form of the potential:
\bea{V_\sigma&=-\frac{1}{2}{\bar \mu}^2\sigma^2+\frac{1}{4}\lambda\sigma^4-\frac{1}{6}\lambda_6\sigma^6\\
V_{SB}&=-\frac{1}{4}\bar{m} \epsilon\sigma \, \mbox{tr} \left[U + U^\dag\right]\ .
}
The second and third lines contain the mass and kinetic terms of vector mesons, respectively.
\begin{widetext}
\noindent
We also rewrite the nucleon part of the Lagrangian as
\begin{align}
\lag_{N}=&{\bar\psi_{1r}}i\gamma^\mu{{D}_\mu}\psi_{1r}
+{\bar\psi_{1l}}i\gamma^\mu{{D}_\mu}\psi_{1l}
+{\bar\psi_{2r}}i\gamma^\mu{{D}_\mu}\psi_{2r}
+{\bar\psi_{2l}}i\gamma^\mu{{D}_\mu}\psi_{2l}\notag\\
&-m_0\left[{\bar\psi_{1l}}\psi_{2r}-{\bar\psi_{1r}}\psi_{2l}
-{\bar\psi_{2l}}\psi_{1r}+{\bar\psi_{2r}}\psi_{1l}\right]\notag\\
&-g_1\sigma\left[{\bar\psi_{1r}} U^\dag \psi_{1l}+{\bar\psi_{1l}} U\psi_{1r}\right]
-g_2\sigma\left[{\bar\psi_{2r}} U \psi_{2l} +{\bar\psi_{2l}} U^\dag \psi_{2r}\right]\notag\\
&-a_{\rho NN}\left[{\bar\psi_{1l}} {\gamma^\mu}
(\xi_L^\dag{\hat\alpha_{\parallel\mu}}\xi_L)\psi_{1l}
+{\bar\psi_{1r}} \gamma^\mu
(\xi_R^\dag{\hat\alpha_{\parallel\mu}}\xi_R) \psi_{1r}\right]\notag\\
&-a_{\rho NN}\left[{\bar\psi_{2l}} {\gamma^\mu}
(\xi_R^\dag{\hat\alpha_{\parallel\mu}}\xi_R)\psi_{2l}
+{\bar\psi_{2r}} \gamma^\mu
(\xi_L^\dag{\hat\alpha_{\parallel\mu}}\xi_L) \psi_{2r}\right]\notag\\
&-a_{0 NN} \, \mbox{tr} \left[ {\hat\alpha_{\parallel\mu}} \right]\,
\left( \bar{\psi_{1l}} \gamma^\mu \psi_{1l} +{\bar\psi_{1r}} \gamma^\mu \psi_{1r}
+ {\bar\psi_{2l}} {\gamma^\mu} \psi_{2l} +{\bar\psi_{2r}} \gamma^\mu \psi_{2r}\right)
\ .
\end{align}
\end{widetext}

The vacuum expectation value (VEV) of the $\sigma$ field, denoted by $\sigma_0$, is determined by the stationary condition for the potential, as we
explained
above.
The non-zero $\sigma_0$ breaks the chiral symmetry spontaneously and generates the masses of nucleons as
\begin{align}
{\cal L}_{mass}=-
\begin{pmatrix}
\bar\psi_1&\bar\psi_2
\end{pmatrix}
\begin{pmatrix}
g_1\sigma_0&m_0\gamma_5 \\ -m_0\gamma_5&g_2\sigma_0
\end{pmatrix}
\begin{pmatrix}
\psi_1 \\ \psi_2
\end{pmatrix}
\ .
\end{align}
We obtain the masses of the positive-parity and negative-parity nucleons by diagonalizing the  mass matrix.
Here, we write the mass eigenstates as $N_+$ and $N_-$, which are related to $\psi_1$ and $\psi_2$ as
\begin{align}
\begin{pmatrix}
N_+\\ N_-
\end{pmatrix}
=
\begin{pmatrix}
\cos\theta & \gamma_5\sin\theta\\
-\gamma_5\sin\theta & \cos\theta
\end{pmatrix}
\begin{pmatrix}
\psi_1\\ \psi_2
\end{pmatrix}\ ,\label{eq:lotd}
\end{align}
where $\theta$ is the mixing angle given by
\begin{align}
\tan2\theta=\frac{2m_0}{(g_1+g_2)\sigma_0}\ \label{eq:mix_angle}.
\end{align}
The mass eigenvalues are determined as
\begin{align}
m_\pm=\frac{1}{2}\left(\sqrt{(g_1+g_2)^2\sigma_0^2+4m_0^2}\mp(g_1-g_2)\sigma_0\right)\ ,
\end{align}
where $m_+$ and $m_-$ are the masses of positive and negative parity baryons, respectively.%
~\footnote{
As we will explain in the next section, we use $m_+ = 939\,$MeV and $m_- = 1535\,$MeV as inputs to determine the values of $g_1$ and $g_2$ for given $m_0$.
When we take $m_+$ as the mass of the negative parity baryon and $m_-$ as the one of positive parity baryon, i.e. $m_- = 939\,$MeV and $m_+ = 1535\,$MeV, the determined values of $g_1$ and $g_2$ are exchanged.  One can verify that this exchange does not cause any physical difference by swapping $\psi_1$ with $\psi_2$.  }
From this expression, one can easily see that the spontaneous chiral symmetry breaking is responsible for the mass differences of the
parity partners.

\section{Determination of model parameters}
\label{parameterF}

In this section, we determine the 10 unknown parameters in this model by performing
a global fit with chosen $m_0$ to masses and the pion decay constant in free space,
and to normal nuclear matter properties. In this global fitting we have used $m_0=900, 800, 700, 600$ and $500$ MeV
as we have found that with $m_0=400$ MeV or less we cannot reproduce the normal nuclear matter properties such as
incompressibility.

Determined parameters are summarized in Table \ref{parametersummary}.
Now, we describe how we fix the model parameters and what are the inputs.

\begin{table}[h]
\begin{center}
\caption{Determined model parameters for given $m_0$. Here $m_\omega=783$ MeV, $m_\rho=776$ MeV and ${\bar
m}\epsilon=m_\pi^2f_\pi$.}\label{parametersummary}
\begin{tabular}{c||c|c|c|c|c}
\hline
$m_0$[MeV]& 500 &600&700&800&900\\
\hline\hline
$g_1$&\ 15.4 \ &\ 14.8 \ &\ 14.2 \ &\ 13.3 \ &\ 12.3 \ \\
$g_2$&\ 8.96 \ &\ 8.43 \ &\ 7.76 \ &\ 6.94 \ &\ 5.92 \ \\
$g_{\omega NN}$&\ 11.4 \ &\ 9.12 \ &\ 7.31 \ &\ 5.67 \ &\ 3.54 \ \\
$g_{\rho NN}$&\ 8.05 \ &\ 6.97\ &\ 7.46 \ &\ 7.75 \ &\ 8.75 \ \\
$\bar{\mu}$[MeV]&\ 435&\ 434 \ &\ 402 \ &\ 316 \ &\ 109 \ \\
$\lambda$&\ 40.5 \ &\ 39.4 \ &\ 34.5 \ &\ 22.5 \ &\ 4.26 \ \\
$\lambda_6$&\ 16.3 \ &\ 15.4 \ &\ 13.5 \ &\ 8.66 \ &\ 0.607 \ \\
\hline\hline
\end{tabular}
\end{center}
\end{table}

First, we determine 6 parameters by using the physical inputs  listed in Table \ref{vacinput}.
We choose the mass of positive- (negative-) parity nucleons as $m_+=939$ MeV ($m_-=1535$ MeV).
As in some literatures, one may also try $m_-=1200$ MeV, but we have chosen the lightest and observed one.
The pion mass and decay constant are $m_\pi=140$ MeV and $f_\pi=93$ MeV.

\begin{table}[h]
\begin{center}
\caption{Physical inputs in vacuum (MeV).}\label{vacinput}
\begin{tabular}{c|c|c|c|c|c}
\hline
$m_+$ & $m_-$ &$m_\omega$ & $m_\rho$ & $f_\pi$ &$m_\pi$ \\
\hline\hline
\ 939\ &\ 1535\ &\ 783\ &\ 776\ &\ 93\ &\ 140\ \\
\hline\hline
\end{tabular}
\end{center}
\end{table}

Next, the remaining parameters are fixed by
the saturation density, the binding energy, the incompressibility and the symmetry energy for the normal nuclear matter at zero temperature.
The normal nuclear matter properties for the fit are discussed below, and  empirical values of them are summarized in Table~\ref{nuclearproperty}.

The typical value of the nuclear saturation density and binding energy are $0.16$ fm$^{-3}$ and $-16$ MeV, respectively.
 With $m_+= 939$ MeV we have

\bea{\rho(\mu_B^*=923~{\rm MeV} )=\rho_0&=0.16\,{\rm fm^{-3}}\\
\brac{\frac{E}{A}-m_+}_{\rho_0}=\brac{\frac{\epsilon}{\rho_B}-m_+}_{\rho_0}&=-16~{\rm MeV}\ .
}
We can consider  another property of nuclear matter, the incompressibility, which is given by the curvature of binding energy at saturation density
and corresponds to the ``hardness'' of the
matter:
\bea{
K=\left.9\rho_0^2\frac{\partial^2(E/A)}{\partial\rho^2}\right|_{\rho_0}=\left.9\rho_0\frac{\partial\mu_B}{\partial\rho}\right|_{\rho_0}
\ .
}
The symmetry energy per nucleon is from the difference of proton and neutron and given as
\bea{E_{sym}(\rho_B)&=\frac{1}{2!}\frac{\partial^2 (E/A)}
{\partial \delta^2}\notag\\
&=\frac{1}{2!}\frac{\partial^2 (\epsilon/\rho)}{\partial \delta^2}
\ ,
}
where the asymmetry parameter $\delta$ is defined as
\bea{\delta\equiv\frac{\rho_p-\rho_n}{\rho_B}={\frac{2\rho_I}{\rho_B}}\ .
}

We fit the remaining four-parameters
 to the empirical values shown in Table \ref{nuclearproperty}.

\begin{table}[h]
\begin{center}
\caption{Physical inputs. $\mu_B^*=923$MeV. We took the value of incompressibility and symmetry energy from \cite{Shlomo2006} and \cite{lattimer_lim} respectively.} \label{nuclearproperty}
\begin{tabular}{c|c|c|c}
\hline
$\rho_0(\mu_B^*)$[fm$^{-3}$]&$E/A(\mu_B^*)-m_+$[MeV]&$K$[MeV]&$E_{sym}$[MeV]\\
\hline\hline
0.16&-16&240&31\\
\hline\hline
\end{tabular}
\end{center}
\end{table}
We would like to stress that we can reproduce the  value of the incompressibility
by the inclusion of the six-point interaction of the scalar meson $\sigma$, in contrast to
the previous analyses by parity doublet models~\cite{Zschiesche:2006zj,Sasaki:2010bp} where
it seems difficult to reproduce the small value of the incompressibility.

\section{Equation of State}\label{EoS}
In this section, we
study the  equation of state for
cold nuclear matter
using the model constructed in the previous sections.

\begin{figure}[h]
\begin{center}
{\includegraphics[width=7.0cm]{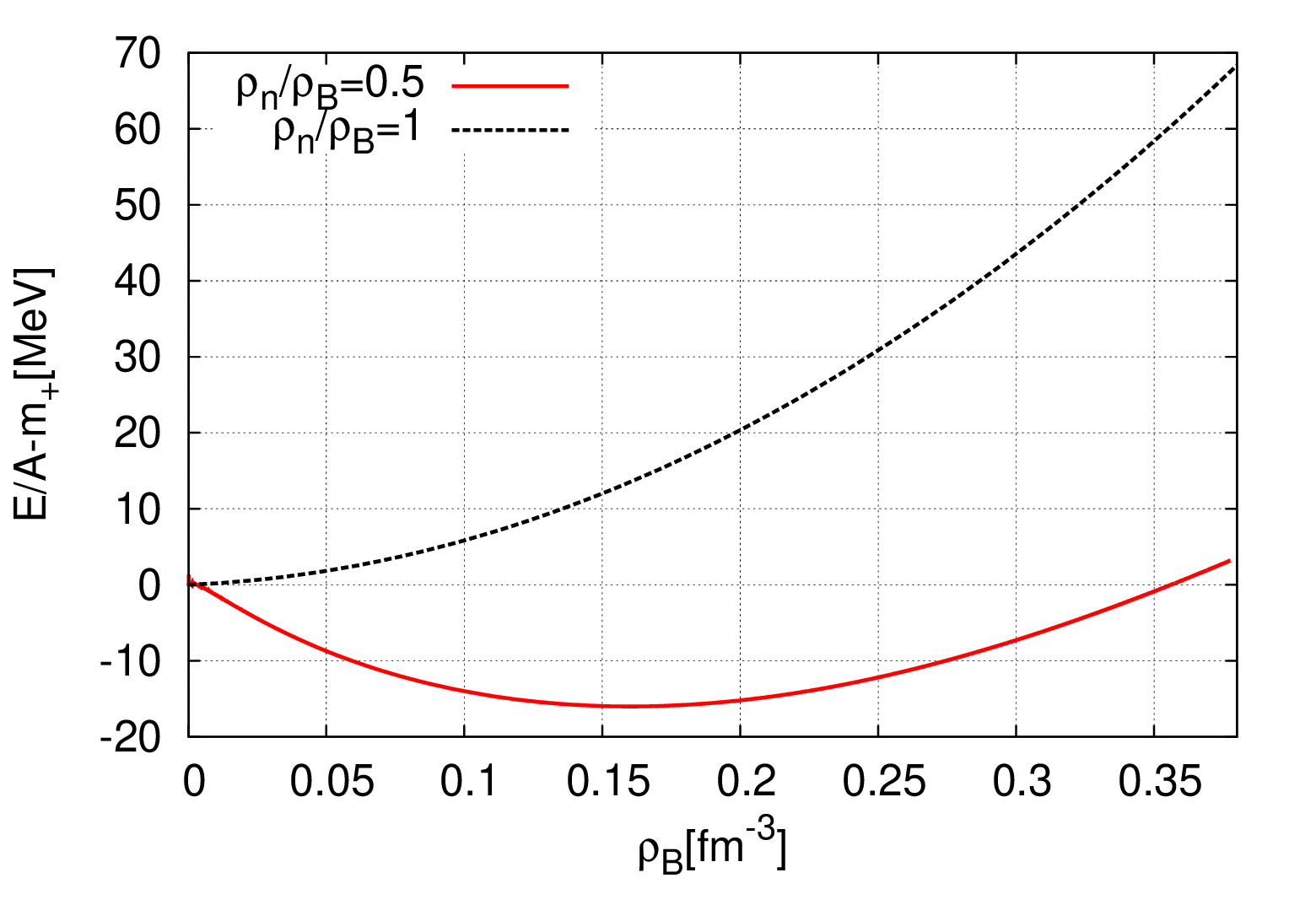}}
\\
{\includegraphics[width=7.0cm]{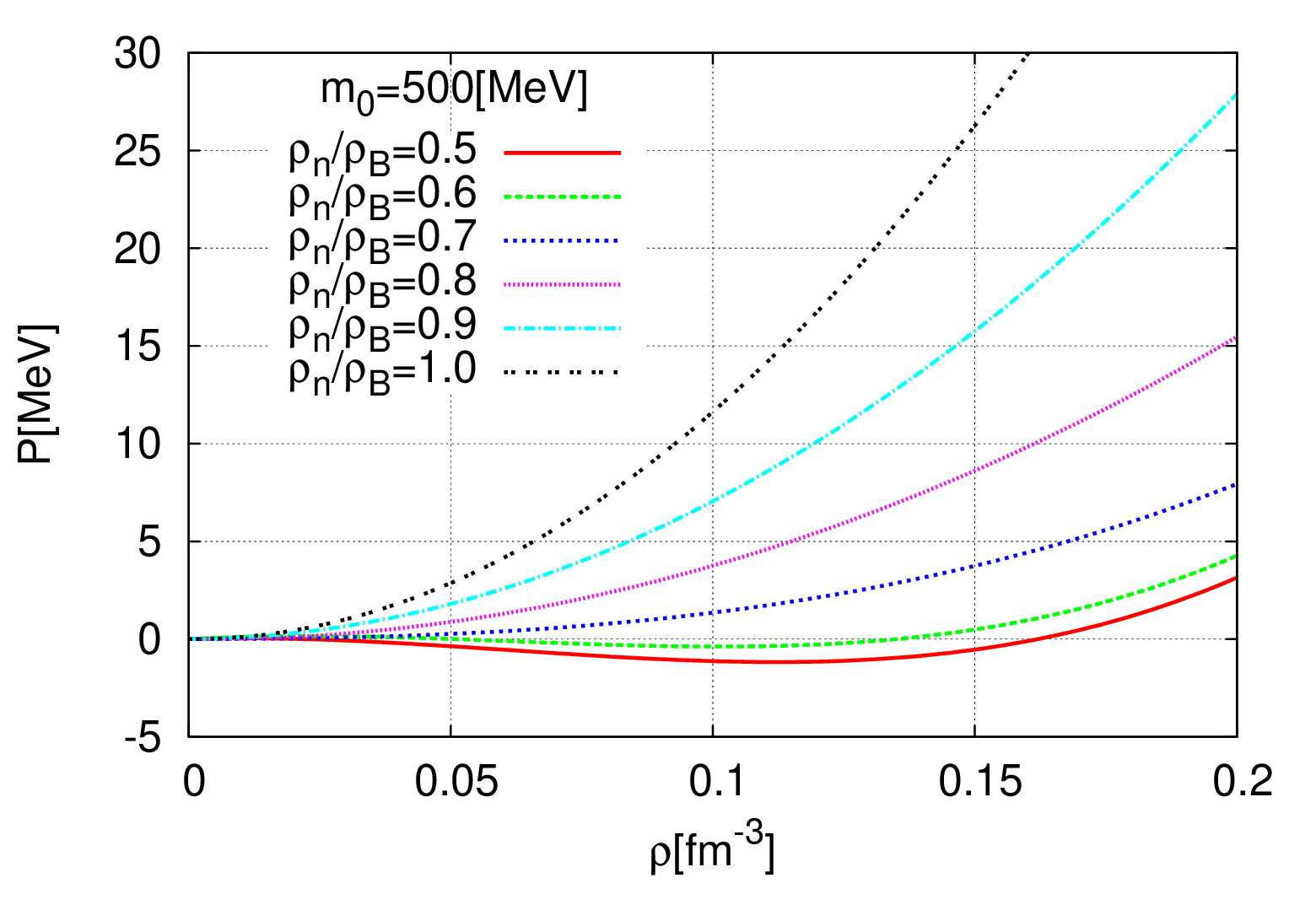}}
\end{center}
\caption{
Density dependence of the binding energy (upper panel) and the pressure (lower panel) for
$m_0=500$ MeV.
 $\rho_B$ is the baryon number density and $\rho_n$ is the neutron number density:
$\rho_n/\rho_B = 0.5$ implies symmetric nuclear matter and $\rho_n/\rho_B = 1$  pure neutron matter.
}
\label{asymbindene_asympress}
\end{figure}

  In Fig.~\ref{asymbindene_asympress},
we show the dependence of the binding energy (upper panel) and
the pressure (lower panel) on the baryon number density for
$m_0=500$ MeV as an example.
The red line
 in the upper panel of Fig.~\ref{asymbindene_asympress} shows the dependence of the binding energy on the baryon number density of
 symmetric matter.
 From this, one can easily see that the binding energy is actually minimized at $\rho_B = \rho_0$, which implies that the existence of the bound nuclei and
 the saturation property are actually reproduced.%
\footnote{ We would like to stress that we do not use the minimization condition as an input. }

The lower panel shows that the pressure for $\rho_n/\rho_B=0.5$ is negative in the low density region, which means that the liquid phase of hadron coexists
with the gas phase.
In  asymmetric matter,
on the other hand,
when the degree of asymmetry, $\rho_n/\rho_B$, is around 0.6$\sim$0.7, the coexistence phase disappears and the pressure increases monotonically for larger
asymmetry, which implies that
the bound nuclei do not exist for $\rho_n/\rho_B > 0.7$.%
~\footnote{ Actually, the critical value of $\rho_n/\rho_B$ depends on the chiral invariant mass: $\rho_n/\rho_B\sim0.7$ for $m_0=500$ MeV and $\rho_n/\rho_B\sim0.8$ for $m_0=900$ MeV.}
In case of the pure neutron matter ($\rho_n/\rho_B=1$), the energy density as well as the pressure monotonically increases with  the baryon number
density.
 The EoS (pressure) of asymmetric nuclear matter is also discussed in molecular dynamics \cite{Lopez:2013cza} and many body perturbation
\cite{Wellenhofer:2015qba}. Though our analysis is done within the mean field approximation, the qualitative behaviors in symmetric and neutron matter
from \cite{Lopez:2013cza, Wellenhofer:2015qba} are similar to our result.

Next, we study the slope parameter, $L$, which is the gradient of the symmetry energy at the saturation density:
\bea{L=3\rho_0\left.\frac{d E_{sym}(\rho_B)}{d\rho_B}\right|_{\rho_B=\rho_0}\ .}
The constraint for the value of $L$ is obtained from some experiments, for example, heavy ion collision experiments, nuclear masses and so
on~\cite{lattimer_lim,slopeparametervalue}.
In Table~\ref{slope}, we list our calculated values of the slope parameters for several choices of the chiral invariant mass $m_0$.
This shows that
the slope parameter hardly depend on $m_0$, all of which are within the allowed region shown in Refs.~\cite{lattimer_lim,slopeparametervalue}.

\begin{table}[h]
\begin{center}
\caption{
Predicted values of the slope parameter
for $m_0 = 500$\,MeV to $900$\,MeV.}
\label{slope}
\begin{tabular}{c||c}
\hline
$m_0$[MeV]&$L$[MeV]\\
\hline\hline
$900$&$75$\\
\hline
$800$&$74$\\
\hline
$700$&$78$\\
\hline
$600$&$78$\\
\hline
$500$&$75$\\
\hline\hline
\end{tabular}
\end{center}
\end{table}

\begin{figure}[h]
\begin{center}
{\includegraphics[width=7.0cm]{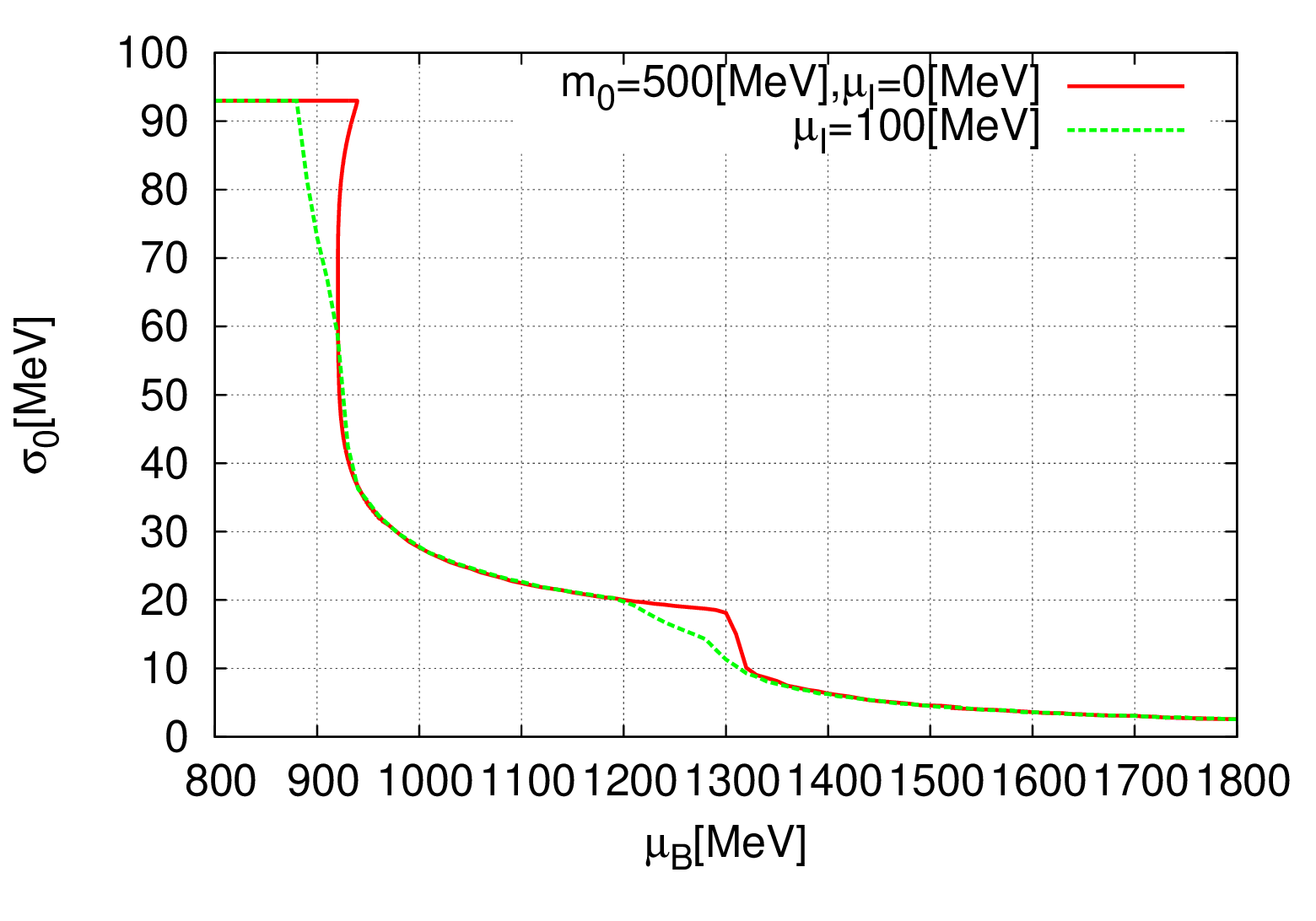}}
\\
{\includegraphics[width=7.0cm]{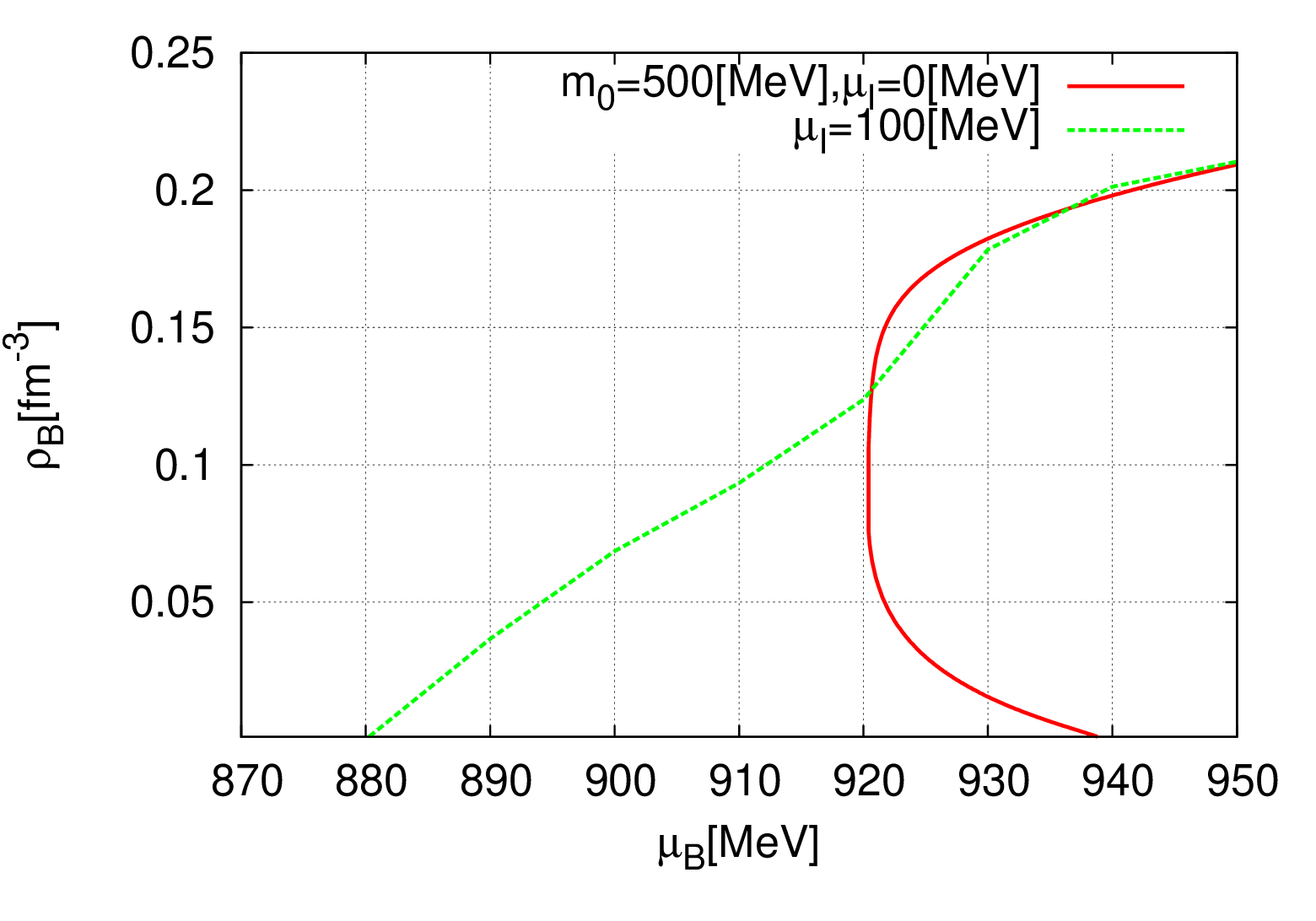}}
\end{center}
\caption{
Density dependence of $\sigma_0$ (upper panel) and chemical potential dependence of $\rho_B$ (lower panel) for
$m_0=500$ MeV and $\mu_I$=0.
}
\label{mubvssigma0_mubvsrhob}
\end{figure}

In Fig.~\ref{mubvssigma0_mubvsrhob},
we show the dependence of the VEV of the $\sigma$ field (upper panel) and the baryon number density $\rho_B$ (lower panel) on the baryon number chemical potential $\mu_B$ for $m_0=500$\,MeV in symmetric matter (red curves) and in the pure neutron matter (green curves).

 The upper panel shows that there are two points,  $\mu_B\sim900$\,MeV and $\sim1300$\,MeV, where the value of $\sigma_0$ changes rapidly.
In the case of  symmetric matter shown by the red curve, the first jump for $\mu_B\sim900$\,MeV can be identified with the first order phase transition from
the liquid phase to the gas phase as the baryon number density (an order parameter of the liquid-gas transition) undergoes sudden change around $\mu_B\sim
900$ MeV (see the lower panel of Fig.~\ref{mubvssigma0_mubvsrhob}).%
~\footnote{One may schematically understand why the chiral condensate drops at the liquid-phase transition, where the baryon number density jumps as we increase
the baryon chemical potential, through the Pauli exclusion principle.
As the baryon density increases, the low-lying phase
space relevant for quark-antiquark condensates is occupied by the fermions (quarks in nucleons in this case) that constitute the Fermi sea; therefore,
forming quark-antiquark condensates requires much energy. At the liquid-gas transition point, there is a sudden increase in the number density, and so
we could expect that the chiral (quark-antiquark) condensate changes drastically.  }
Then, the phase transition around $\mu_B\sim 1300$ MeV can be naturally identified as  the chiral phase transition.
In  symmetric nuclear matter (red dashed-line), the liquid-gas phase transition is first-order and there exists coexistence phase.
The existence of coexistence phase in the nuclear liquid-gas phase transition has been confirmed in the experiments~ \cite{liquid-gastransitionexperiment}.
The coexistence phase in symmetric matter disappears by increasing $\mu_I$, and the liquid-gas transition becomes second order as suggested
in~\cite{Muller:1995ji}.

\begin{figure}[h]
\begin{center}
{\includegraphics[width=7.0cm]{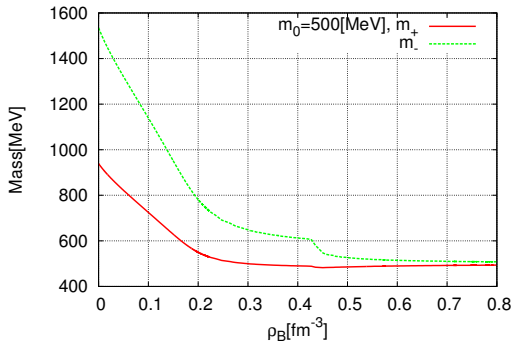}}
\end{center}
\caption{Density dependence of the
effective nucleon masses for $m_0=500$ MeV at $\mu_I=0$. }
\label{massatm0500}
\end{figure}
Figure \ref{massatm0500} shows  the density dependence of the effective mass of positive (negative) parity nucleon, $m_+$ ($m_-$).
As $\rho_B$ increases,  $m_+$ and $m_-$ gradually get close to the chiral invariant mass.
This is a feature from parity doublet structure, and two nucleon masses degenerate to $m_0$ when the chiral symmetry is completely restored.
However, the mass difference between the parity partners remains finite in our model due to the current quark mass.
The critical density for chiral symmetry restoration depends also on the chiral invariant mass, which will be discussed in detail in the next section.

\section{Phase Diagram}\label{phaseD}
In this section, we explore the phase structure of our model at finite temperature and density with the isospin asymmetry.
Our primary interest here is to see how the onset of the chiral and  liquid-gas phase transition depends on the isospin asymmetry and
the chiral invariant nucleon mass $m_0$.
In this study, we will not consider the charged pion condensation, and so we take $|\mu_I|<m_\pi$.

\begin{figure}[h]
\begin{center}
{\includegraphics[width=7.0cm]{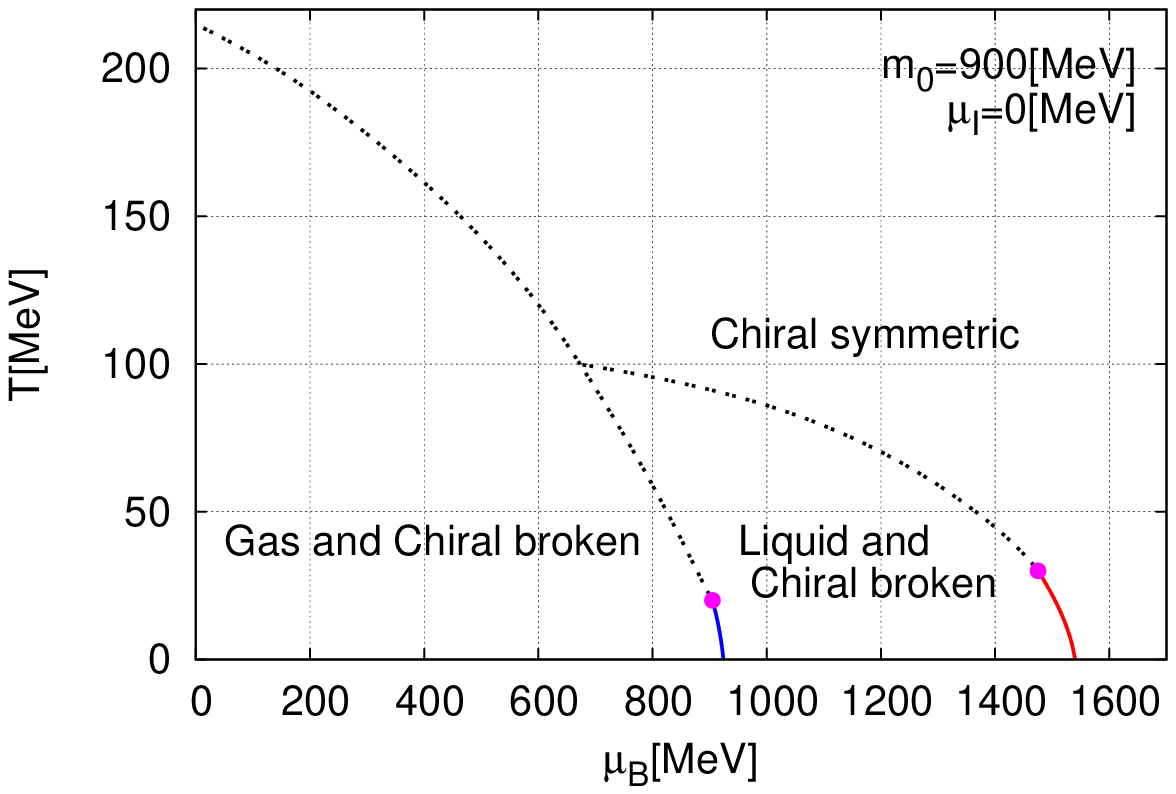}}
\\
{\includegraphics[width=7.0cm]{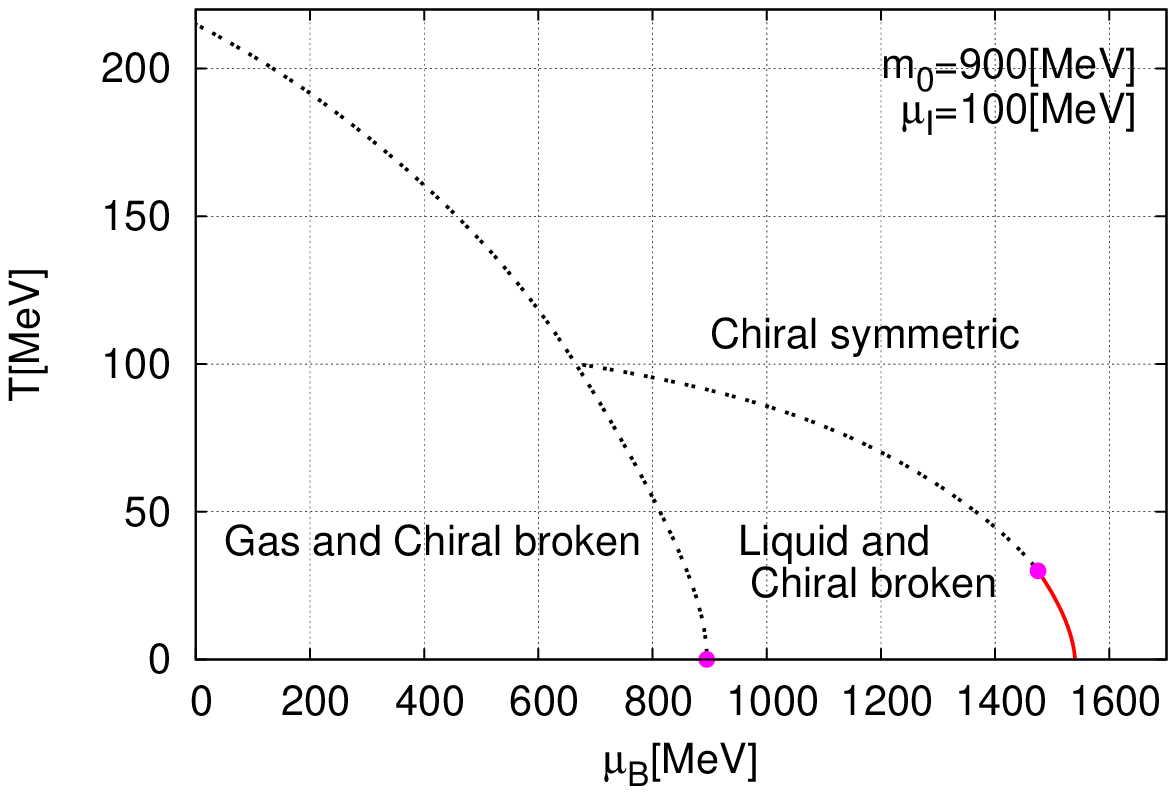}}
\end{center}
\caption{
Phase diagrams for $m_0=900$ MeV at $\mu_I=0$ (upper panel) and $\mu_I = 100$\,MeV (lower panel).  The solid line is for the first order phase transition, the dashed line for the crossover, and the point for the critical point (2nd order).}
\label{phasediagrams_at_m0900}
\end{figure}

The upper panel of Figure \ref{phasediagrams_at_m0900} corresponds to the phase diagram at $\mu_I=0$, where we have
the first order liquid-gas phase transition (blue solid line) around $\mu_B=900$\,MeV and chiral phase transition (red solid line) around $\mu_B=1500$\,MeV
at zero and finite but small temperatures.
The magenta dots are second order critical points, and green dashed line shows the crossover.

The two crossover lines meet at a point around $\braa{\mu_B,~ T}=\braa{700~{\rm MeV}, 100~{\rm MeV}}$ as seen in a study with a parity doublet model~\cite{Sasaki:2010bp}.

The lower panel of Figure~\ref{phasediagrams_at_m0900} shows the phase diagram of asymmetric matter with $m_0=900$\,MeV at $\mu_I=100$ MeV.
The trend of the liquid-gas and chiral phase transition is almost the same with the upper panel ($\mu_I=0$); however, the liquid-gas transition becomes second
order even at zero temperature as briefly mentioned in the previous section.

\begin{figure}[h]
\begin{center}
{\includegraphics[width=7.0cm]{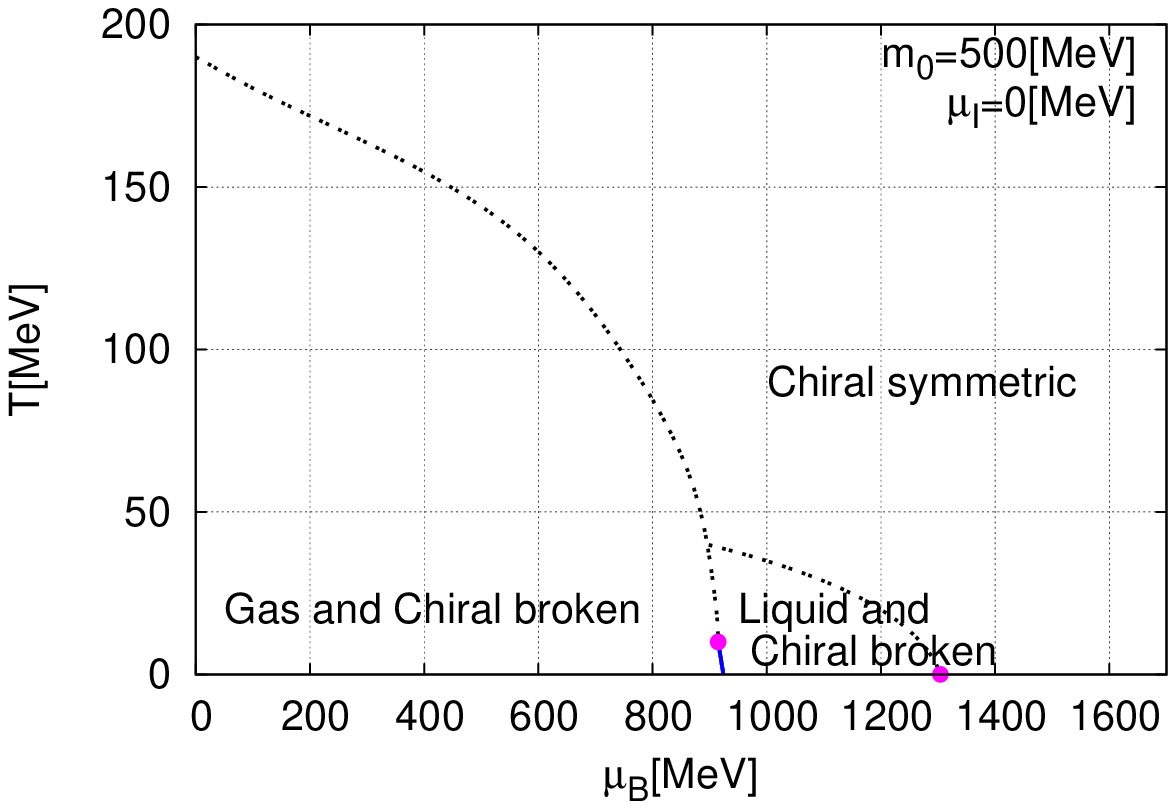}}
\\
{\includegraphics[width=7.0cm]{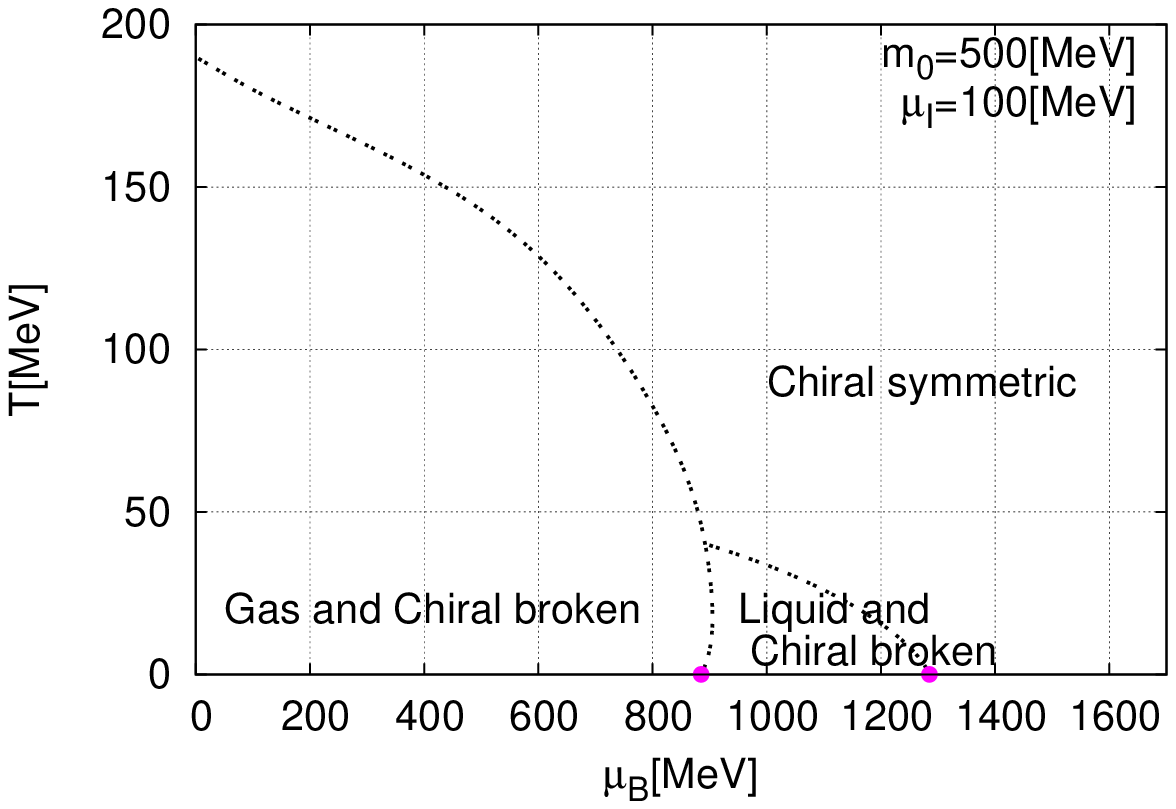}}
\end{center}
\caption{The phase diagrams for $m_0=500$ MeV.}
\label{phasediagrams_at_m0500}
\end{figure}

Now, we study how the chiral invariant mass affects the phase diagram. We take $m_0=500$ MeV as an example.
Since the chiral invariant mass measures the amount of spontaneous chiral symmetry breaking needed for the nucleon masses and their mass splitting,
we may expect that it mainly affects chiral symmetry in the phase diagram.
The results are shown in  Fig.~\ref{phasediagrams_at_m0500}.
As expected, the nature of the liquid-gas transitions does not change, while the first order chiral phase transition becomes second order.
Furthermore, as $m_0$ decreases, the critical density (chemical potential) for chiral transition decreases monotonically as summarized in Table
\ref{criticalmubmui0}.
Note that the saturation density is an input, and so the corresponding chemical potential, in our study. This is why
the critical density (chemical potential) for liquid-gas phase transition does not change.

\begin{table}[h]
\begin{center}
\caption{Critical values of the baryon chemical potential ${\mu_B}^c$ at $T=0$ and $\mu_I=0$. }
\label{criticalmubmui0}
\begin{tabular}{c||cc||cc}
\toprule
$m_0$[MeV]&${\mu_B}^c_{lg}$[MeV] & ${\rho_B}^c_{lg}$[fm$^{-3}$]
&${\mu_B}^c_{\chi}$[MeV]&${\rho_B}^c_{\chi}$[fm$^{-3}$]\\ \cline{2-3}
\hline\hline
$900$&923&0.16&1540&2.98\\
\hline
$800$&923&0.16&1643&1.75\\
\hline
$700$&923&0.16&1554&1.09\\
\hline
$600$&923&0.16&1426&0.69\\
\hline
$500$&923&0.16&1305&0.44\\
\hline\hline
\end{tabular}
\end{center}
\end{table}
When we employ $m_0=900$ MeV, the critical density for chiral transition is about $17\rho_0$.
On the other hand, the critical density becomes $\sim 3\rho_0$ with $m_0=500$ MeV.
Comparing with a previous study in a parity doublet model with $m_0=800$ MeV~\cite{Sasaki:2010bp}, we find that our result for
the symmetric matter is almost the same with the one from the previous study.
Note, however, that we can explore the phase diagrams with the chiral invariant mass of the range $500~{\rm MeV}<m_0<900{\rm MeV}$ as we introduced
the six-point interaction.

Table \ref{criticalmubmui100} shows the critical baryon chemical potential and density for $\mu_I=100$ MeV;
comparing with Table \ref{criticalmubmui0}, we observe that the critical values become (slightly) smaller.
 Here the liquid-gas phase transition is second order so that ${\rho_B}_{lg}^c$ must be $0$.

\begin{table}[h]
\begin{center}
\caption{Critical values of the baryon chemical potential and density at $T=0$ and $\mu_I=100$ MeV.}
\label{criticalmubmui100}
\begin{tabular}{c||cc||cc}
\toprule
$m_0$[MeV]&${\mu_B}^c_{lg}$[MeV] &${\rho_B}^c_{lg}$[fm$^{-3}$]
&${\mu_B}^c_{\chi}$[MeV]&${\rho_B}^c_{\chi}$[fm$^{-3}$]\\ \cline{2-3}
\hline\hline
$900$&891&0.0&1537&2.98\\
\hline
$800$&891&0.0&1637&1.74\\
\hline
$700$&891&0.0&1543&1.09\\
\hline
$600$&888&0.0&1410&0.68\\
\hline
$500$&881&0.0&1285&0.43\\
\hline\hline
\end{tabular}
\end{center}
\end{table}

\section{Summary and Discussion}\label{SD}

We have constructed a model for asymmetric nuclear matter by extending the parity doublet model. We introduced vector mesons ($\rho$ and $\omega$)
through hidden local symmetry and also included the six-point interaction of of $\sigma$ meson.
We fixed our model parameters with chosen $m_0$ by performing a global fit to physical inputs
(masses and pion decay constant in free space and nuclear matter properties).
With the six-point potential, we were able to reproduce
normal nuclear matter properties with $m_0$ in the range from $500~{\rm MeV}$ to $900~{\rm MeV}$.

We first studied the equation of state and the phase diagram of dense symmetric matter at finite temperature.
We observed that the slope parameter at the saturation density satisfies the constraint from heavy ion experiments and neutron star observations,
 for instance, see \cite{lattimer_lim, slopeparametervalue} and found that the chiral condensate changes drastically at the chiral and liquid-gas transition
 points.

 We then moved on to asymmetric dense matter with non-zero iso-spin chemical potential.
We showed that the first order nature of the liquid-gas transition disappears in asymmetric matter and
the critical densities for the chiral transition becomes smaller with increasing iso-spin chemical potentials,
which are in agreement with the results from existing literatures.
We also showed that smaller chiral invariant nucleon mass favors smaller critical density for chiral phase transition both in symmetric and asymmetric dense matter.

In this study we don't consider interesting phenomena in dense (asymmetric) matter such as transition from nuclear matter to hyperonic matter,
charged pion condensation with large iso-spin chemical potential, and so on. These will be relegated to future works.

Finally, we discuss the chiral invariant nucleon mass.
In our work, we chose the chiral invariant mass $m_0=500$-$900$\,MeV to reproduce the properties of normal nuclear matter.
However, there are other choices from various studies.
In \cite{Zschiesche:2006zj, Sasaki:2010bp}, nuclear matter was studied in a parity doublet model and rather large value of the
chiral invariant mass was used $m_0\sim 800$ MeV.
On the other hand, in \cite{jidookahosaka} the authors determined $m_0$  from the decay width of $N^*\to N\pi$ to be $m_0=270$ MeV, while in
\cite{Gallas:2009qp} they used the decay modes of $N^*\to N\pi$ and $a_1\to\pi\gamma$ to obtain $m_0\sim500$ MeV.
Recently, the study of parity doublet structure using lattice QCD~\cite{latticeparitydoublet} shows that the positive parity nucleon mass changes very small
near the deconfinement transition, which may imply that  $m_0$ does exist in nature and  its value is close to the positive parity
nucleon mass.
The existence of chiral invariant mass is also discussed in the context of Skyrmion crystal~\cite{dongkuoleemachleidtrho}.
Since the values of the chiral invariant mass are diverse in literatures, it is quite important and interesting to narrow down the value of $m_0$ and
to dig further down to the role of $m_0$ in hadron physics.

\section*{Acknowledgments}

The work of Y.M. is supported in part by the Nagoya University Program for Leading Graduate Schools funded by the Ministry of Education of the Japanese Government under the program number N01.
The work of Y.K. was supported  by the Rare Isotope Science Project of Institute for Basic Science funded by Ministry of Science, ICT and Future Planning and
National Research Foundation of Korea (2013M7A1A1075766).
M. H. was supported in part by the JSPS Grant-in-Aid for Scientific Research (S) No.22224003 and (c)No. 24540266.

%%%%%%%%%%%%%%%%%%%%%%%%%%%%%%%%%%%%%%%%%%%%%

\end{document}